# Dispersion-managed dark solitons in erbium-doped fiber lasers

**Han Zhang[1], Dingyuan Tang[1]*, Mustapha Tlidi[2], Luming Zhao and Xuan Wu**

[1]School of Electrical and Electronic Engineering, Nanyang Technological University, 639798, Singapore

[2]Optique Nonlinéaire Théorique, Université libre de Bruxelles (U.L.B.), CP.231, Campus Plaine, B-1050 Bruxelles, Belgium

*: edytang@ntu.edu.sg, corresponding author.

We report on the observation of dispersion-managed (DM) dark soliton emission in a net-normal dispersion erbium-doped fiber laser. We found experimentally that dispersion management could not only reduce the pump threshold for the dark soliton formation in a fiber laser, but also stabilize the single dark soliton evolution in the cavity. Numerical simulations have also confirmed the DM dark soliton formation in a fiber laser.

In the past three decades soliton formation in single mode fibers (SMFs) has been extensively investigated [1]. It is now well-recognized that the dynamics of the solitons is governed by the nonlinear Schrödinger equation (NLSE). In particular, in the anomalous dispersion SMFs bright solitons can be formed, while in the normal dispersion SMFs dark solitons, characterized as a localized intensity dip on a continuous wave (CW) background [2], can be formed. Experimentally, bright and dark soliton formation and propagation in SMFs were reported [1-7]. Bright and dark solitons of the NLSE type have also been observed in SMF lasers [8, 9]. Despite of the fact that pulse propagation in a fiber laser is not exactly the same as in the SMFs, e. g. a pulse circulating in a laser cavity also subjects to the actions of the laser gain and cavity output, it was found that under appropriate laser operation conditions the dynamics of the formed solitons could still be well described by the NLSE.

A feature of fiber lasers is that their cavities can be easily dispersion managed (DM), meaning that the cavity is made of fibers with opposite sign of group velocity dispersion. Light circulation in a DM cavity fiber laser is analog to the light propagation in an endless DM fiber transmission line [10]. It has been theoretically shown that such a DM fiber transmission line supports both bright and dark solitons, known as the DM bright or dark solitons [11-13]. Experimentally, the DM bright solitons have been observed in DM cavity fiber lasers [14]. However, to the best of our knowledge, no experimental observation on the DM dark solitons has been reported. In this paper, we report on the first experimental observation of DM dark solitons in a net normal dispersion erbium-doped fiber laser. We found experimentally that the formation of DM dark soliton has lower pump threshold as compared with the dark soliton formation in an all-normal-dispersion fiber laser. In addition, the DM dark soliton operation of a fiber laser is less sensitive to the environmental perturbations. Numerical simulations have also confirmed DM dark soliton formation in our fiber laser.

A dispersion-managed ring cavity fiber laser as shown in Fig. 1 was used. The cavity consists of a piece of ~5.0 m, 2880 ppm Erbium-doped fiber (EDF) with a group velocity dispersion (GVD) parameter of -32 (ps/nm)/km and ~8.0 m single mode fiber (SMF) with a GVD parameter of 18 (ps/nm)/km. The rest of fibers used are the dispersion compensation fiber (DCF) with a GVD parameter of -2 (ps/nm)/km. The total cavity length is ~18.2 m. A polarization dependent isolator was used to force the unidirectional operation of the cavity, and an in-line polarization controller (PC) was inserted in the cavity to fine-tune the birefringence of the cavity. A 50% fiber coupler was used to output the signal. The laser was pumped by a high power Fiber Raman Laser source (KPS-BT2-RFL-1480-60-FA) of wavelength 1480 nm. The maximum pump power can reach to 5 W. An optical spectrum analyzer (Ando AQ-6315B) and a 350 MHZ oscilloscope (Agilen 54641A) together with a 2 GHZ photo-detector were simultaneously used to monitor the laser operation.

The total dispersion of the cavity is estimated ~0.3224 $ps^2$. We note that the laser has exactly the same cavity configuration as a nonlinear polarization rotation (NPR) mode locking fiber laser [9]. Indeed, depending on the linear cavity phase delay bias (LCPDB) setting, self-started mode-locking could be obtained in the laser. A typical mode locked state of the laser is shown in Fig. 2. The mode locked pulses exhibit the DM dissipative soliton features. The central wavelength of the solitons is at 1585.7 nm, and their spectrum has a 3 dB bandwidth of ~13.6 nm. The FWHM DM dissipative soliton pulse width is ~266 fs. Therefore, their time-bandwidth product (TBP) is ~0.452.

Starting from a bright dissipative soliton operation state, we then fixed all the other laser parameters but continuously tuned the orientation of the PC. It was observed that the laser emission changed from the bright soliton emission first to a CW operation state, and then to a dark soliton emission state, as illustrated in Fig. 3. In Fig. 3a we have shown both the CW laser emission spectrum and the dark pulse emission spectrum. Under the dark pulse emission, the spectrum becomes obviously broader. We note that the dark pulse emission state is not a mode locked state of the laser, compared with Fig. 2, the LCPDB is now set in the non-mode-locking regime [9]. On the oscilloscope trace, a dark soliton pulse is identified by a narrow intensity dip on the strong CW background. Depending on the pump strength, either multiple or single dark solitons could be obtained. In the case of multiple dark solitons, the solitons have different darkness, indicating that different from the bright solitons, no soliton energy quantization effect exists. The FWHM width of the dark pulse is narrower than 500 ps, which is below the resolution of our detection system. Unfortunately, due to the low repetition rate of the dark pulses, the actual dark pulse width cannot be measured with the conventional autocorrelation technique either. Based on the 3 dB spectral bandwidth of ~0.8 nm, as shown in Fig. 3a, we estimated that the pulse width is around ~3 ps, if a transform-limited hyperbolic-tangent profile is assumed.

We have experimentally investigated the effect of the anomalous dispersion SMF on the DM dark soliton by deliberately changing the SMF length. The net cavity dispersion then changed from all-normal to net-anomalous. When SMF was completely removed, and the laser has an all-normal dispersion short cavity, very high pump power, normally higher than 2.5 W, was required to obtain dark pulse emission. In this case it is difficult to obtain stable single dark soliton. New dark solitons were constantly automatically formed [9]. After SMF was spliced into the cavity and its length was increased, it was observed that the pump threshold for the dark soliton formation progressively decreased, and single dark soliton became less difficult to obtain. It seemed that the insertion of the anomalous dispersion SMF could stabilize the dark soliton propagation in the laser cavity. Under an appropriate cavity DM strength and pump strength, stable single dark soliton emission states have been obtained. We found experimentally that the 3 dB spectral bandwidth of the single dark soliton could not be broader than ~0.9 nm. Beyond it, the single dark soliton became unstable, and eventually a new dark soliton appeared. The laser displayed a clearly different multiple dark soliton formation scenario than an all-normal dispersion cavity fiber laser. When the anomalous dispersion SMF was elongated to ~10 m and thereby the net cavity dispersion shifted to the net-anomalous dispersion regime, no dark soliton was observed.

We further numerically simulated the DM dark soliton formation in our fiber laser. We used a model as described in [15]. To make the simulations comparable with our experiment, we used the following parameters: the orientation of the intra-cavity polarizer to the fiber fast birefringent axis $\Phi$=0.13 $\pi$; the nonlinear fiber coefficient $\gamma$=3 $W^{-1}km^{-1}$; the erbium fiber gain bandwidth $\Omega_g$ =24 nm; fiber dispersions $D''_{EDF}$= -32 (ps/nm) /km, $D''_{SMF}$= 18 (ps/nm) /km, $D''_{DCF}$= -2 (ps/nm) /km and $D'''$=0.1 ($ps^2$/nm)/km; cavity linear birefringence $L/L_b$=0.01 and the nonlinear gain saturation energy $P_{sat}$=500 pJ, $L_{EDF}$=5 m and $L_{DCF}$=5.2 m, and the length of SMF was changed from 0 m to 10 m, therefore, the total cavity dispersion was varied from 0.2215 $ps^2$ to -0.0125 $ps^2$.

In all our simulations an arbitrary weak dip input was used as the initial condition. It was found that when the LCPDB was biased in the non-mode-locking regime, stable dark soliton could always be obtained. In the case of an all-normal-dispersion cavity, we found that through progressively increasing the laser gain, a dark soliton can hardly break into multiple dark solitons. Increasing the pump strength increased the CW level, and an initially black soliton gradually became a gray soliton, and eventually vanished in the noisy CW background. However, if anomalous dispersion SMF was introduced, not only the formation threshold of the dark solitons became much lower, but also the formed DM dark solitons exhibited different features. As the pump strength is increased, the DM dark solitons split. Fig. 4 shows the numerically calculated parameter domain for stable single DM dark soliton of our laser. Depending on the net cavity dispersion, the range of gain variation for stable single DM dark soliton is limited, e.g. in a cavity with net-normal dispersion of ~0.0437 $ps^2$, where ~7.6 m SMF was used, gain could not exceed an upper threshold ~580 $km^{-1}$. Otherwise, the DM dark soliton would break automatically.

Bound states of dark solitons were also numerically obtained, as shown in Fig. 5. In the simulation we have used net-dispersion 0.0343 $ps^2$. Under a low laser gain of ~ 485 $km^{-1}$, very stable single DM dark soliton with pulse width of ~3.7 ps and a modulation depth of ~0.92, which indicates a nearly black DM soliton, was obtained. When the laser gain was raised to 510 $km^{-1}$, a new dark soliton was formed, as shown in Fig. 5b. Like the experimental observations, the dark solitons have different darkness. The separation between the dark solitons is ~3.8 ps, representing the formation of a DM black-grey soliton pair. It is to point out that although bound DM dark soliton pair was numerically obtained, it remains a challenge to experimentally observe it. The main difficulty is how to experimentally measure two closely spaced dark pulses. A streak camera might be necessary to visualize it.

In conclusion, we have first experimentally observed DM dark solitons in a net-normal dispersion erbium-doped fiber laser. We found that in a DM cavity not only the formation threshold of a dark soliton is lower, as compared with that of an all-normal dispersion cavity, but also the state of stable single dark soliton could be easily obtained. DM dark soliton formation in our laser was also numerically simulated. Moreover, numerical simulations have revealed a bound state of black-grey solitons. Our study shows that dark soliton emission is an intrinsic feature of the normal dispersion fiber lasers.

## References:

1. L. F. Mollenauer, R. H. Stolen, and J. P. Gordon, “Experimental Observation of Picosecond Pulse Narrowing and Solitons in Optical Fibers,” Phys. Rev. Lett. **45**, 1095 (1980).
2. Y. S. Kivshar and G. P. Agrawal, Optical solitons: from fibers to photonic crystals (Academic, 2003).
3. P. Emplit, J. P. Hamaide, F. Reynaud, and A. Barthelemy, “Picosecond steps and dark pulses through nonlinear single mode fibers,” Opt. Commun. **62**, 374 (1987).
4. A. M. Weiner, J. P. Heritage, R. J. Hawkins, R. N. Thurston, E. M. Kirschner, D. E. Leaird, and W. J. Tomlinson, “Experimental Observation of the Fundamental Dark Soliton in Optical Fibers,” Phys. Rev. Lett. **61**, 2445 (1988).
5. G. Millot, E. Seve, S. Wabnitz and M. Haelterman, “Dark-soliton-like pulse-train generation from induced modulational polarization instability in a birefringent fiber,” Opt. Lett. **23**, 511 (1998).
6. T. Sylvestre, S. Coen, P. Emplit, and M. Haelterman, “Self-induced modulational instability laser revisited: normal dispersion and dark-pulse train generation,” Opt. Lett. **27**, 482 (2002).
7. M. Tlidi and L. Gelens, “High-order dispersion stabilizes dark dissipative solitons in all-fiber cavities,” Opt. Lett. **35**, 306 (2009).
8. I. N. Duling III, “All-fiber ring soliton laser mode locked with a nonlinear mirror,” Opt. Lett. **16**, 539-541 (1991).

9. H. Zhang, D. Y. Tang, L. M. Zhao, and X. Wu, “Dark pulse emission of a fiber laser,” Phys. Rev. A **80**, 045803 (2009).
10. V. S. Grigoryan, T. Yu, E. A. Golovchenko, C. R. Menyuk, and A. N. Pilipetskii, “Dispersion-managed soliton dynamics,” Opt. Lett. **22**, 1609 (1997).
11. Y. J. Chen, “Dark solitons in dispersion compensated fiber transmission systems,” Opt. Commun. 161, 267 (1999).
12. M. J. Ablowitz and Z. H. Musslimani, “Dark and gray strong dispersion-managed solitons,” Phys. Rev. E **67**, R 025601 (2003).
13. M. Stratmann and F. Mitschke, “Chains of temporal dark solitons in dispersion-managed fiber,” Phys. Rev. E **72**, 066616 (2005).
14. R.-M. Mu, V. S. Grigoryan, C. R. Menyuk, G. M. Carter, and J. M. Jacob, “Comparison of theory and experiment for dispersion-managed solitons in a recirculating fiber loop,” IEEE J. Sel. Topics Quant. Electron., **6**, 248–257, (2000).
15. D. Y. Tang, H. Zhang, L. M. Zhao, and X. Wu, “Observation of High-Order Polarization-Locked Vector Solitons in a Fiber Laser,” Phys. Rev. Lett. **101**,153904 (2008).

## References:

1. L. F. Mollenauer, R. H. Stolen, and J. P. Gordon, Phys. Rev. Lett. **45**, 1095 (1980).
2. Y. S. Kivshar and G. P. Agrawal, Optical solitons: from fibers to photonic crystals (Academic, 2003).
3. P. Emplit, J. P. Hamaide, F. Reynaud, and A. Barthelemy, Opt. Commun. **62**, 374 (1987).
4. A. M. Weiner, J. P. Heritage, R. J. Hawkins, R. N. Thurston, E. M. Kirschner, D. E. Leaird, and W. J. Tomlinson, Phys. Rev. Lett. **61**, 2445 (1988).
5. G. Millot, E. Seve, S. Wabnitz and M. Haelterman, Opt. Lett. **23**, 511 (1998).
6. T. Sylvestre, S. Coen, P. Emplit, and M. Haelterman, Opt. Lett. **27**, 482 (2002).
7. M. Tlidi and L. Gelens, Opt. Lett. **35**, 306 (2009).
8. I. N. Duling III, Opt. Lett. **16**, 539-541 (1991).
9. H. Zhang, D. Y. Tang, L. M. Zhao, and X. Wu, Phys. Rev. A **80**, 045803 (2009).
10. V. S. Grigoryan, T. Yu, E. A. Golovchenko, C. R. Menyuk, and A. N. Pilipetskii, Opt. Lett. **22**, 1609 (1997).
11. Y. J. Chen, Opt. Commun. **161**, 267 (1999).
12. M. J. Ablowitz and Z. H. Musslimani, Phys. Rev. E **67**, R 025601 (2003).
13. M. Stratmann and F. Mitschke, Phys. Rev. E **72**, 066616 (2005).
14. V. S. Grigoryan and C. R. Menyuk, Opt. Lett. **23**, 609 (1998).

15. D. Y. Tang, H. Zhang, L. M. Zhao, and X. Wu, Phys. Rev. Lett. **101**,153904 (2008).

## Figure captions:

Fig.1: Schematic of the vector dark soliton fiber laser. WDM: wavelength division multiplexer. EDF: erbium doped fiber. SMF: single mode fiber. PDI: polarization dependent isolator. PCs: polarization controllers.

Fig. 2: Optical spectrum of DM bright soliton. Insert: its autocorrelation trace and oscilloscope trace.

Fig. 3: (a) Spectra of DM dark soliton and CW. (b) Oscilloscope traces, upper: single dark soliton; down: multiple dark solitons.

Fig. 4: Region of existence of DM dark solitons in the (Dispersion, Gain) plane.

Fig. 5: Evolution of DM dark solitons in time domain with the net-cavity dispersion ~0.343 $ps^2$: (a) Gain= 485 $km^{-1}$; (b) Gain= 510 $km^{-1}$;

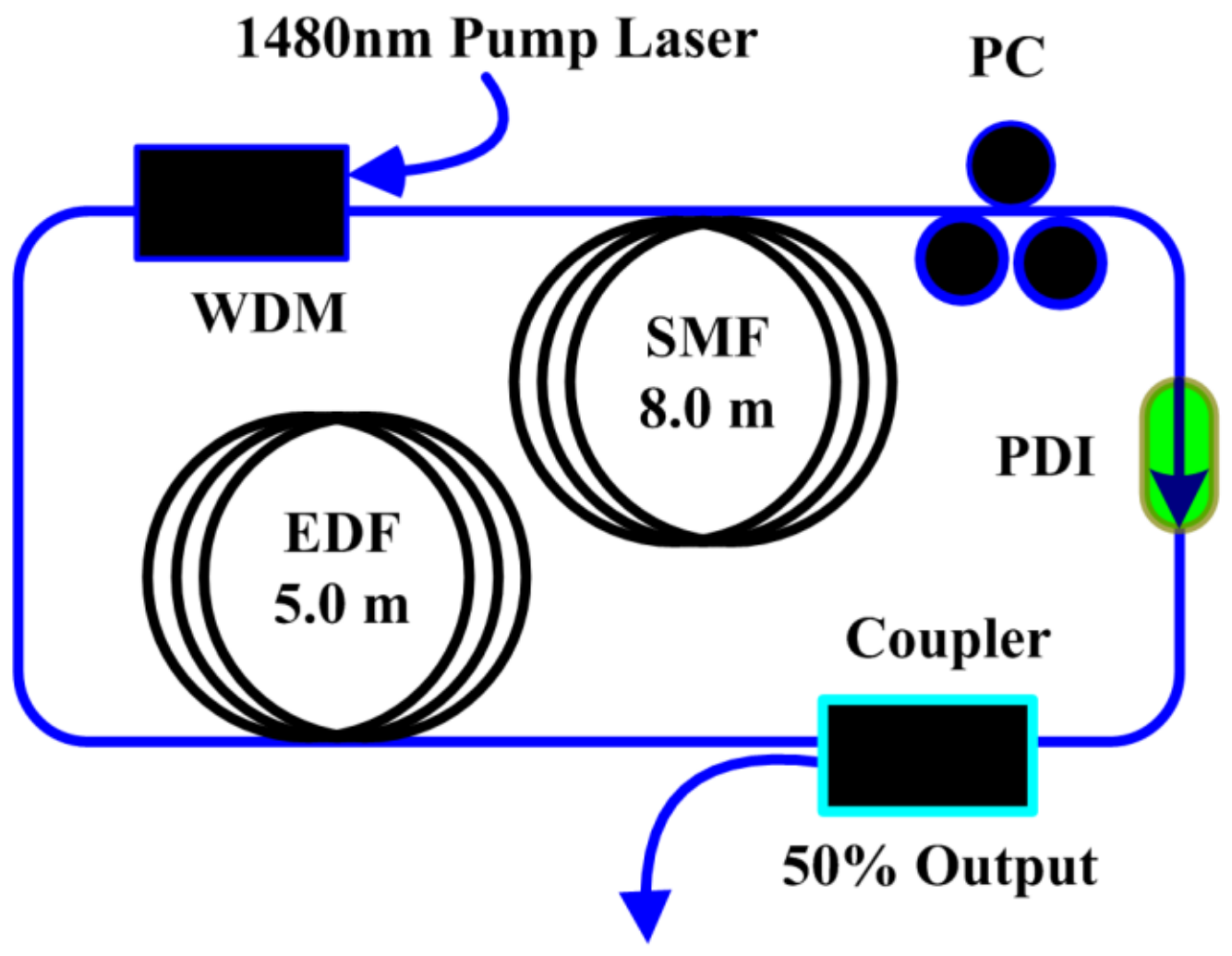

Fig.1 H. Zhang et al.

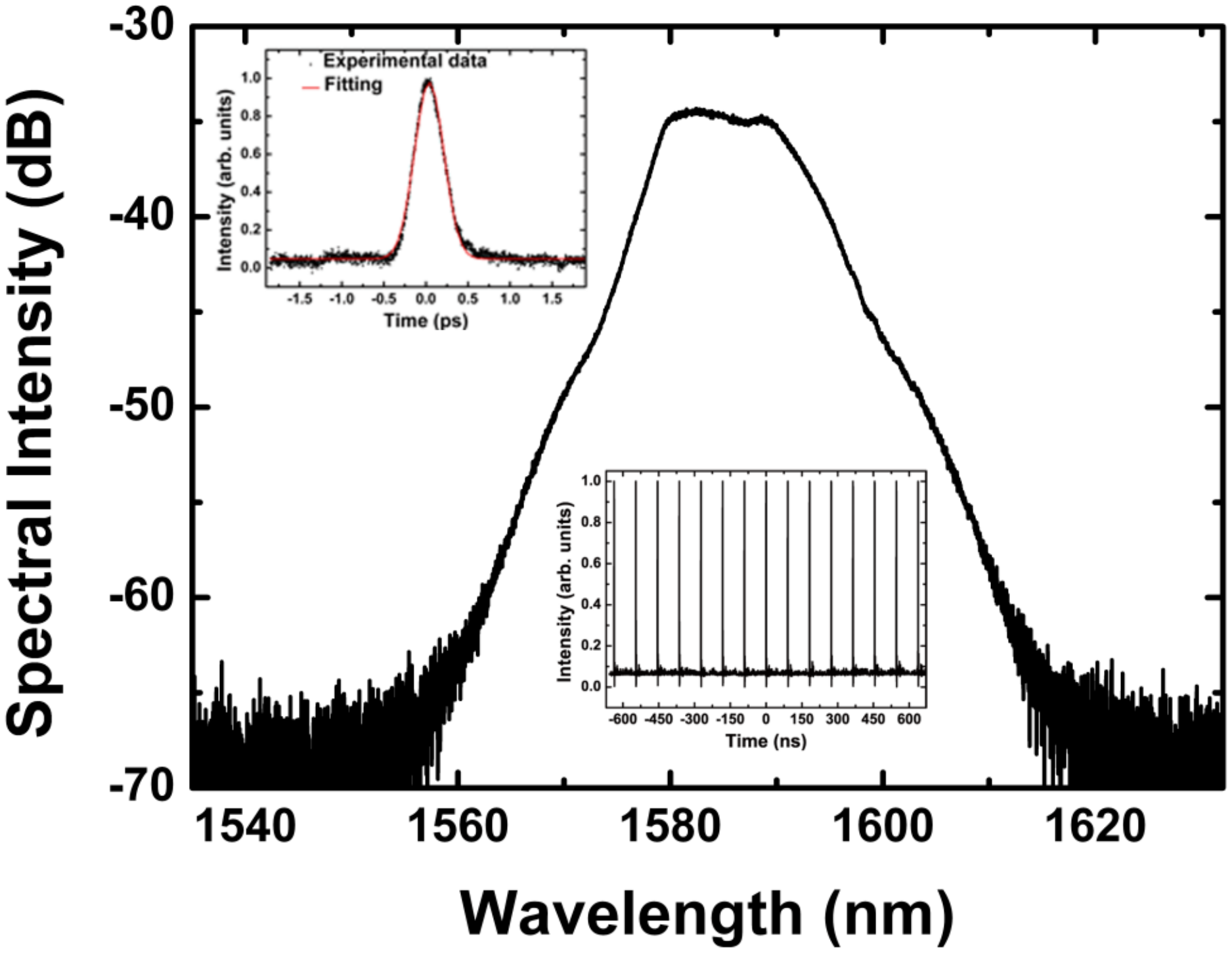

Fig.2 H. Zhang et al.

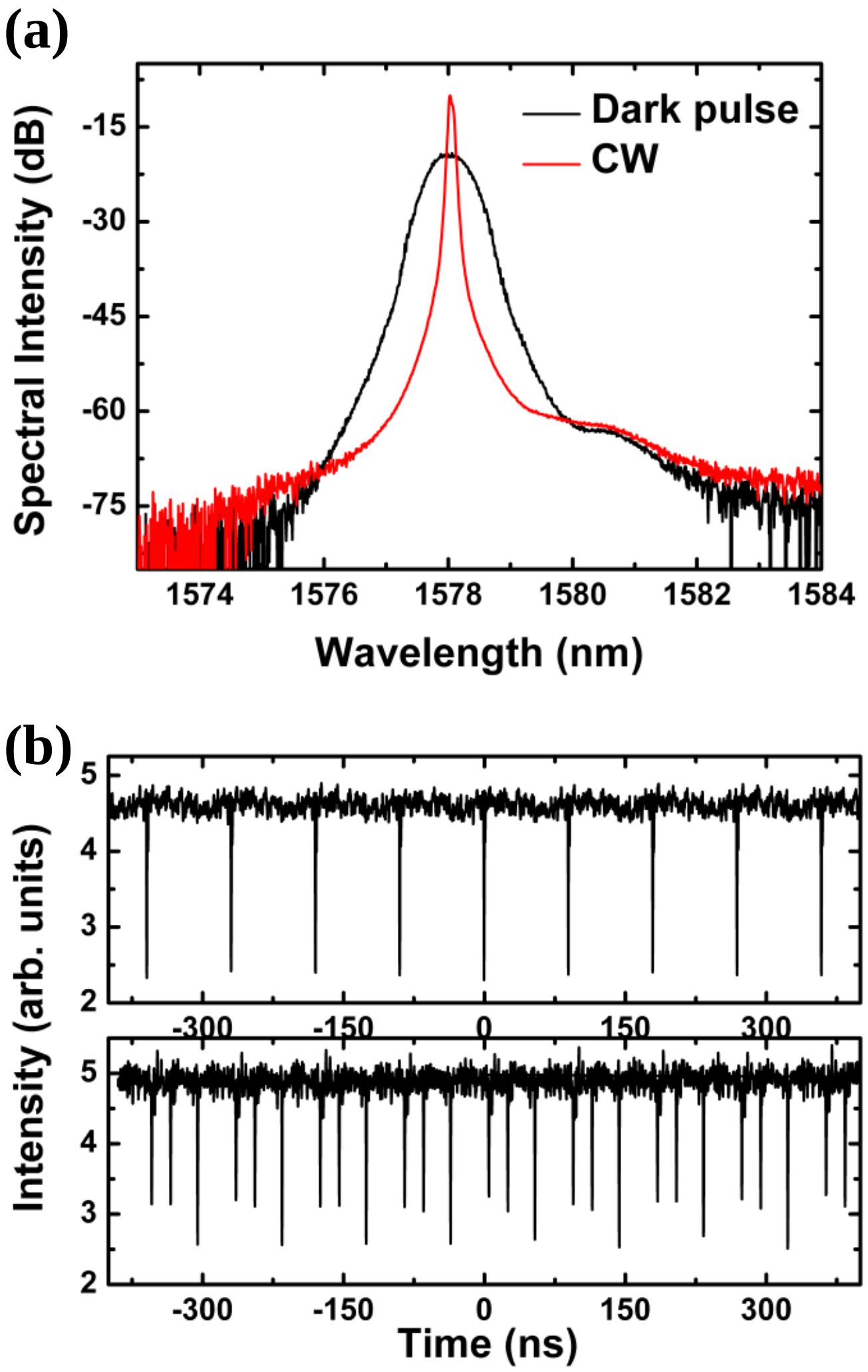

Fig.3 H. Zhang et al.

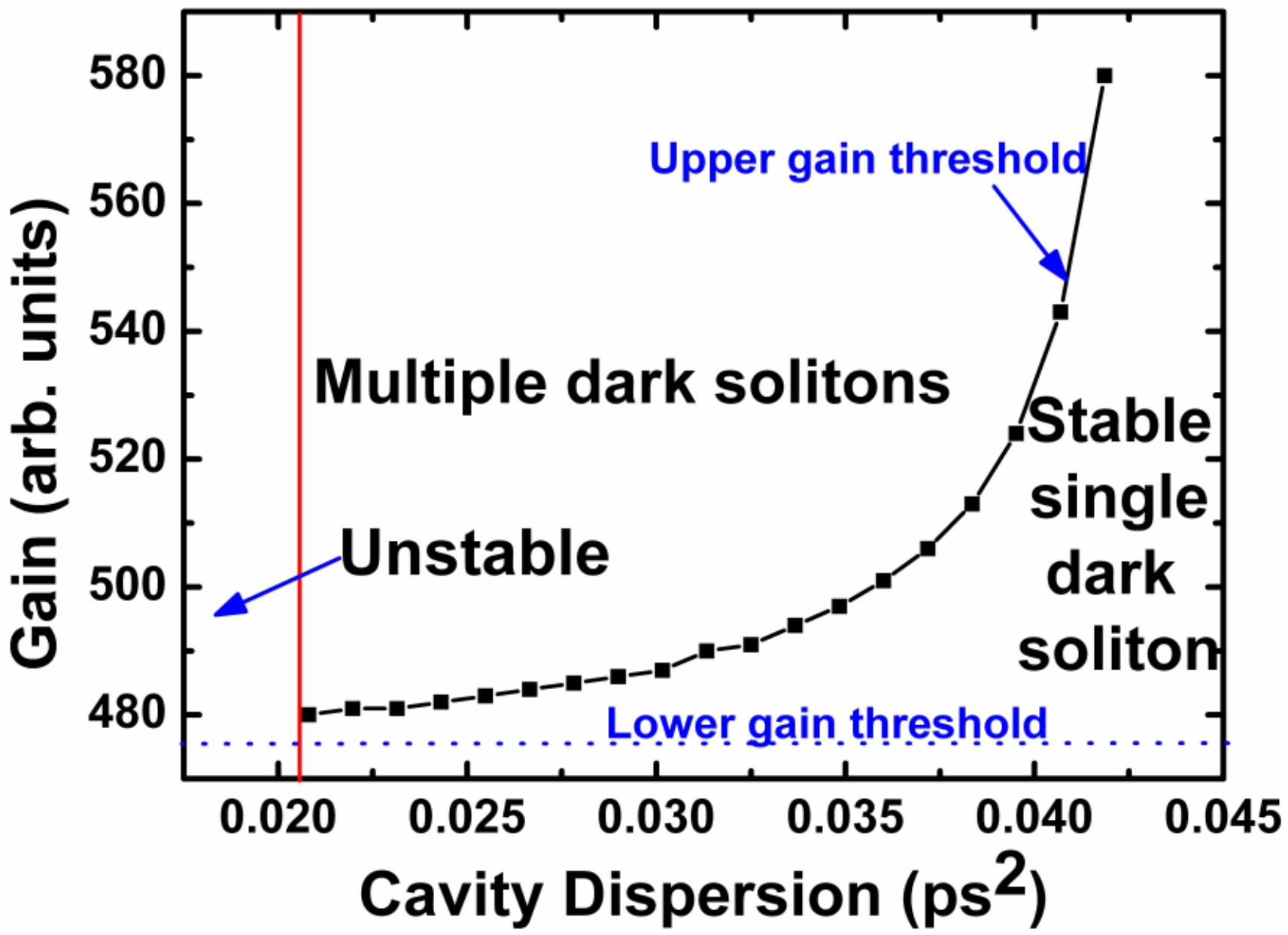

Fig.4: H. Zhang et al.

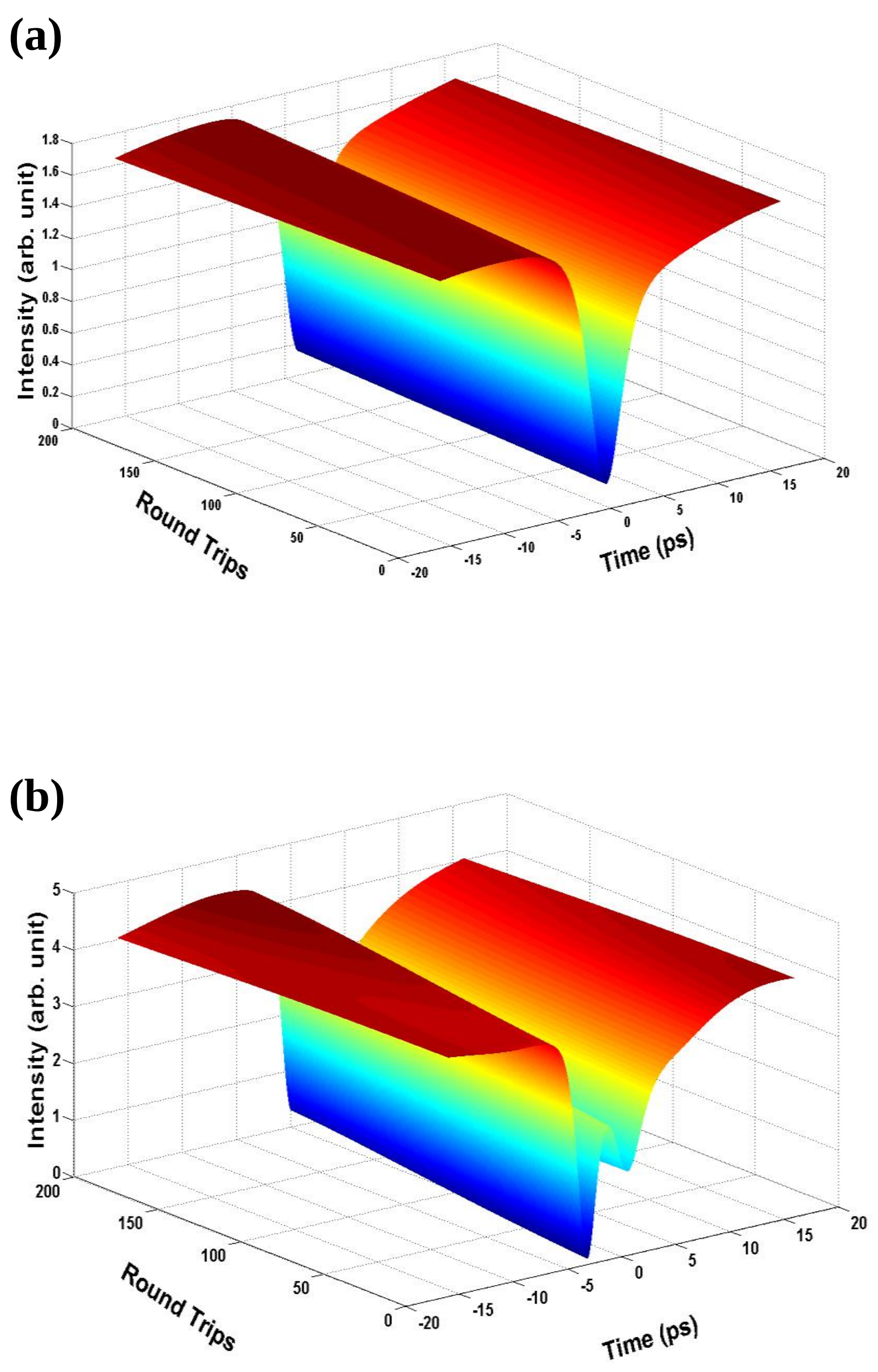

Fig.5: H. Zhang et al.

Simulation model based on the following equations:

$$\frac{\partial u}{\partial z} = i\beta u - \delta\frac{\partial u}{\partial t} - \frac{i\mathrm{D}''}{2}\frac{\partial^2 u}{\partial t^2} + \frac{\mathrm{D}'''}{6}\frac{\partial^3 u}{\partial t^3} + i\gamma\left(|u|^2 + \frac{2}{3}|v|^2\right)u + \frac{i\gamma}{3}v^2u^* + \frac{g}{2}u + \frac{g}{2\Omega_g^2}\frac{\partial^2 u}{\partial t^2}$$

$$\frac{\partial v}{\partial z} = -i\beta v + \delta\frac{\partial v}{\partial t} - \frac{iD''}{2}\frac{\partial^2 v}{\partial t^2} + \frac{D'''}{6}\frac{\partial^3 v}{\partial t^3} + i\gamma\left(|v|^2 + \frac{2}{3}|u|^2\right)v + \frac{i\gamma}{3}u^2v^* + \frac{g}{2}v + \frac{g}{2\Omega_g^2}\frac{\partial^2 v}{\partial t^2}$$

The orientation of the intra-cavity polarizer to the fiber fast birefringent axis $\Phi=0.13\ \pi$; the nonlinear fiber coefficient $\gamma=3\ \ W^{-1}km^{-1}$; the erbium fiber gain bandwidth $\mathbf{\Omega}_g$ =24 nm; fiber dispersions $D''_{EDF}$= -32 (ps/nm) /km, $D''_{SMF}$= 18 (ps/nm) /km, $D''_{DCF}$= -2 (ps/nm) /km and $D'''$=0.1 ($ps^2$/nm)/km; cavity linear birefringence $L/L_b$=0.01 and the nonlinear gain saturation energy $P_{sat}$=500 pJ, $L_{EDF}$=5 m and $L_{DCF}$=5.2 m,  and the length of SMF was changed from 0 m to 10 m, therefore, the total cavity dispersion was varied from 0.2215 $ps^2$ to -0.0125 $ps^2$.